\documentclass[twoside,twocolumn,9pt]{article}
\usepackage{extsizes}
\usepackage[super,sort&compress,comma]{natbib} 
\usepackage[version=3]{mhchem}
\usepackage[left=1.5cm, right=1.5cm, top=1.785cm, bottom=2.0cm]{geometry}
\usepackage{balance}
\usepackage{mathptmx}
\usepackage{sectsty}
\usepackage{graphicx} 
\usepackage{lastpage}
\usepackage{amsmath}
\usepackage{amssymb}
\usepackage[format=plain,justification=justified,singlelinecheck=false,font={stretch=1.125,small,sf},labelfont=bf,labelsep=space]{caption}
\usepackage{float}
\usepackage{fancyhdr}
\usepackage{fnpos}
\usepackage[english]{babel}
\addto{\captionsenglish}{%
  
}
\usepackage{array}
\usepackage{droidsans}
\usepackage{charter}
\usepackage[T1]{fontenc}
\usepackage[usenames,dvipsnames]{xcolor}
\usepackage{setspace}
\usepackage[compact]{titlesec}
\usepackage{hyperref}

\usepackage{epstopdf}

\definecolor{cream}{RGB}{222,217,201}

\begin{document}

\pagestyle{fancy}
\thispagestyle{plain}
\fancypagestyle{plain}{
\renewcommand{\headrulewidth}{0pt}
}

\makeFNbottom
\makeatletter
\renewcommand\LARGE{\@setfontsize\LARGE{15pt}{17}}
\renewcommand\Large{\@setfontsize\Large{12pt}{14}}
\renewcommand\large{\@setfontsize\large{10pt}{12}}
\renewcommand\footnotesize{\@setfontsize\footnotesize{7pt}{10}}
\makeatother

\renewcommand{\thefootnote}{\fnsymbol{footnote}}
\renewcommand\footnoterule{\vspace*{1pt}%
\color{cream}\hrule width 3.5in height 0.4pt \color{black}\vspace*{5pt}} 
\setcounter{secnumdepth}{5}

\makeatletter 
\renewcommand\@biblabel[1]{#1}            
\renewcommand\@makefntext[1]%
{\noindent\makebox[0pt][r]{\@thefnmark\,}#1}
\makeatother 
\renewcommand{\figurename}{\small{Fig.}~}
\sectionfont{\sffamily\Large}
\subsectionfont{\normalsize}
\subsubsectionfont{\bf}
\setstretch{1.125} 
\setlength{\skip\footins}{0.8cm}
\setlength{\footnotesep}{0.25cm}
\setlength{\jot}{10pt}
\titlespacing*{\section}{0pt}{4pt}{4pt}
\titlespacing*{\subsection}{0pt}{15pt}{1pt}

\renewcommand{\headrulewidth}{0pt} 
\renewcommand{\footrulewidth}{0pt}
\setlength{\arrayrulewidth}{1pt}
\setlength{\columnsep}{6.5mm}
\setlength\bibsep{1pt}

\makeatletter 
\newlength{\figrulesep} 
\setlength{\figrulesep}{0.5\textfloatsep} 

\newcommand{\topfigrule}{\vspace*{-1pt}%
\noindent{\color{cream}\rule[-\figrulesep]{\columnwidth}{1.5pt}} }

\newcommand{\botfigrule}{\vspace*{-2pt}%
\noindent{\color{cream}\rule[\figrulesep]{\columnwidth}{1.5pt}} }

\newcommand{\dblfigrule}{\vspace*{-1pt}%
\noindent{\color{cream}\rule[-\figrulesep]{\textwidth}{1.5pt}} }

\makeatother
\twocolumn[
  \begin{@twocolumnfalse}
\vspace{1em}
\sffamily
\begin{tabular}{m{4.5cm} p{13.5cm} }

 & \noindent\LARGE{\textbf{Effect of Sodium Chloride Adsorption on the Surface Premelting of Ice$^\dag$}} \\
\vspace{0.3cm} & \vspace{0.3cm} \\

 & \noindent\large{Margaret L. Berrens,\textit{$^{a}$} Fernanda C. Bononi,\textit{$^{a}$} and Davide Donadio\textit{$^{a \ddag}$}} \\


 & \noindent\normalsize{
We characterise the structural properties of the quasi-liquid layer (QLL) at two low-index ice surfaces in the presence of sodium chloride (Na$^+/$Cl$^-$) ions by molecular dynamics simulations.  We find that  the presence of a high surface density of Na$^+/$Cl$^-$ pairs changes the surface melting behaviour from step-wise to gradual melting. 
The ions lead to an overall increase of the thickness and the disorder of the QLL, and \textcolor{black}{to a low-temperature roughening transition of the air-ice interface}. The local molecular structure of the QLL is similar to that of liquid water, and the differences between the basal and primary prismatic surface are attenuated by the presence of Na$^+/$Cl$^-$ pairs. 
These changes modify the crystal growth rates of different facets and the solvation environment at the surface of sea-water ice with a potential impact on light scattering and environmental chemical reactions.   
}
\end{tabular}

 \end{@twocolumnfalse} \vspace{0.6cm}

]

\renewcommand*\rmdefault{bch}\normalfont\upshape
\rmfamily
\section*{}
\vspace{-1cm}

\footnotetext{\textit{$^{a}$~Department of Chemistry, University of California Davis, Davis, CA, 95616. }}
\footnotetext{\textit{$^{\ddag}$~E-mail: ddonadio@ucdavis.edu}}

\footnotetext{\dag~Electronic Supplementary Information (ESI) available: [details of any supplementary information available should be included here]. See DOI: 10.1039/cXCP00000x/}


\section{Introduction}
Ice covers $10\%$ of the Earth and its surface is an important catalyst for atmospheric chemical reactions, such as the adsorption, photodegradation, and release of trace gases in snow-packs,\cite{tolbert_reaction_1987} the formation of ozone-depleting species in polar stratospheric clouds,\cite{chu_quantum_2003} and chemical reactions in the troposphere.\cite{ravishankara_heterogeneous_1997,mcfall_nitrate_2018,hullar_photodecay_2020} 
From the melting point down to about -70$^o$ C, the surface of ice features a disordered pre-melted layer known as the quasi-liquid layer (QLL). Its thickness depends on the temperature and on the orientation of the ice facets.\cite{slater_surface_2019} The molecular structure of the QLL dictates the morphology of growing ice crystals, it controls the adsorption of molecules at the air-ice interface, and the thermodynamics and kinetics of the chemical reactions involving these molecules.

The thickness, structure and diffusivity of the QLL at pure ice surfaces have been extensively studied by both experiments and computer simulations. 
There are still substantial discrepancies in the estimate of both the onset temperature of surface melting and the thickness of the QLL, which depend on both the experimental conditions and techniques.\cite{shultz_ice_2017,slater_surface_2019,nagata_surface_2019,sanchez_experimental_2017,bluhm_premelting_2002, dosch_glancing-angle_1995,sazaki_quasi-liquid_2012,murata_thermodynamic_2016}   
Molecular dynamics (MD) simulations
provide a consistent picture of very few (two or three) molten (bi)layers, corresponding to a thickness of about 1 nm, up to 2 K below the melting point.\cite{conde_thickness_2008,benet_premelting-induced_2016,kling_structure_2018,qiu_why_2018,llombart_surface_2020} 
Simulations suggest bilayer-by-bilayer ``{\it rounded}" melting transitions that correlate with the trends observed in sum-frequency generation (SFG) spectra.\cite{sanchez_experimental_2017,llombart_rounded_2020}   
Furthermore, MD shows that the underlying crystalline layers induce an ordered template in the QLL structure that persists up to 2 K below the melting temperature. This dynamic ordering marks the major differences between the QLL and bulk water,\cite{kastelowitz_anomalously_2010, watkins_large_2011, wei_sum-frequency_2002,kling_structure_2018} and it affects the solvation properties of the premelting layer.\cite{niblett_ion_2021} 

Whereas the studies of pure-water ice surfaces still leave unanswered questions, even more remain about how the molecular and mesoscale structure of the premelting layer changes in the presence of adsorbed species. 
In natural environments, ice surfaces are exposed to extrinsic species of various kinds, including ions, inorganic, and organic compounds, which impact the structure of the air-ice interface.\cite{kahan_pinch_2014,zimmermann_adsorption_2016} 
Sodium and chloride are the most abundant ions in seawater and can be found naturally in the atmosphere. Sea salt aerosols deposit on environmental snow and ice, and frozen seawater has a layer of brine at its surface containing a high surface density of Na$^+$ and Cl$^-$ ions. 
Glancing-angle Raman spectroscopy suggested that the surface of seawater ice is markedly different from that of freshwater ice.\cite{kahan_pinch_2014}  According to these measurements, the liquid brine formed at the air-ice interface of seawater ice provides a chemical environment similar to that of bulk supercooled water, possibly leading to changes in the rates of photolysis reactions.\cite{kahan_benzene_2010} 
However, the molecular mechanisms that lead to such a stark difference between pristine and briny ice surfaces and the details of the surface of seawater ice 
are still mostly unexplored.

In this work, we investigate the effect of adsorbed sodium and chloride ions on the structure of the premelting layer at low-index surfaces of hexagonal ice I$_h$ by means of classical MD simulations. 
We analyse the molecular structure of the QLL in the presence of Na$^+$ and Cl$^-$ as a function of temperature and surface orientation. We consider the two-bilayer low-index basal (0001) and primary prismatic (10$\bar{1}$0) surfaces, and we model two naturally occurring Na$^+/$Cl$^-$ surface densities of 0.1 NaCl pair/nm$^2$ and 1 NaCl pair/nm$^2$
The former is similar to the surface density at the surface of snow atop of first-year sea ice, while the latter is similar to the surface density of brine atop of sea ice.\cite{krnavek_chemical_2012} 
Our simulations highlight how the ions affect the premelting behaviour compared to pristine ice surfaces. Specifically, the presence of ion pairs at high surface density turns the discrete bilayer-by-bilayer melting, which was observed for pristine ice basal and primary prismatic surfaces, into gradual surface melting.
Furthermore, solvated Na$^+/$Cl$^-$ pairs alter the surface roughness and its fluctuations thus impacting ice crystal growth, the adsorption of other molecules and the chemical processes at the briny ice surface.







\section{Computational Methods}

Studying the premelting of ice surfaces by MD requires the simulation of systems of several thousands of water molecules for time scales \textcolor{black}{of hundreds of nanoseconds}. The need for relatively large models was made clear in former works that address finite size effects in the characteristic behaviour of premelting transitions.\cite{benet_premelting-induced_2016,qiu_why_2018}
For these reasons, we perform classical MD simulations using the TIP4P/2005 empirical rigid-molecule point-charge model.\cite{abascal_general_2005} This model, parameterised on a set of experimental properties of water among which the temperature of maximum density, reproduces qualitatively the phase diagram of molecular ice and water,\cite{sanz_phase_2004,abascal_dipole-quadrupole_2007} and it is among the best in the class of  rigid non-polarisable forcefields at reproducing the thermodynamic properties of water and ice.\cite{vega_simulating_2011} However, it is important to note that TIP4P/2005 has a melting point of 249.5 \(\pm\) 0.1 K \cite{conde_high_2017}, compared to the experimental value of 273.15 K. 
To describe the interactions between the water molecules and the Na$^+$ and Cl$^-$ ions, we employ the Madrid-2019 forcefield,\cite{garcia_fernandez_melting_2006,benavides_potential_2017} which is a point-charge model based on the method of scaled charges. Its parameters are calibrated using hydration numbers, radial distribution functions, and the density of both aqueous solutions and solid NaCl. 
This forcefield has been used to study the seawater/ice interface, the physical properties of seawater, and the effect of NaCl segregation in ice grain boundaries.\cite{conde_molecular_2018, de_almeida_ribeiro_effect_2021}


MD simulations were carried out using the GROMACS 2020.3 program compiled in double-precision.\cite{berendsen_gromacs_1995, van_der_spoel_gromacs_2005} The MD equations of motion were integrated with a time step of 1 fs, which guarantees the conservation of total energy in a microcanonical test run \textcolor{black}{with a total energy drift of approximately 5$x$10$^{-5}$eV/atom per ns.}. 
Long-range electrostatics were computed using the particle-particle particle-mesh Ewald method.\cite{eastwood_p3m3dp-three-dimensional_1984}
An orthorhombic ice \(I_h\) slab model made of 6144 H\(_2\)O molecules in a supercell with 3D periodic boundary conditions was used for simulations. The slab models were \textcolor{black}{constructed} from bulk ice I$_h$ models, in which proton disorder is generated by a Monte Carlo algorithm that minimises the total dipole of the simulation cell. \cite{grishina_structure_2004, matsumoto_genice_2018} 
Bulk ice was equilibrated in the constant pressure canonical ensemble at a set of temperatures from 190 to 240 K with a stride of 10 K. These well-equilibrated models were then cleaved by introducing a vacuum layer of 100 \AA. This thickness of the vacuum layer is sufficient to prevent interactions between the two juxtaposed surfaces through periodic boundary conditions. 
The slabs were equilibrated for 50 ns in the canonical ensemble, enforced by stochastic velocity rescaling\cite{bussi_canonical_2007} with a coupling constant of 0.1 ps. The same thermostat was used to perform production runs in the canonical ensemble. 
For each one of the temperatures considered, we ran production run simulations for 200 ns, \textcolor{black}{and the structural properties reported in the next section (density profiles, surface structure function, and order parameter distributions) are calculated averaging over 2000 frames taken every 1 ps for each trajectory. Uncertainties are computed as the standard deviation over eight 25-ns-long blocks into which each trajectory is split. We have also verified that the differences among the blocks are due to statistical fluctuations rather than a systematic drift, thus proving that our production runs are sufficiently long.
Furthermore, we averaged the structural properties over the two surfaces of the slab, after verifying that the computed properties are the same on each of them within the statistical error.} 


\begin{figure}[thp]
    \centering
    \includegraphics[width=8cm]{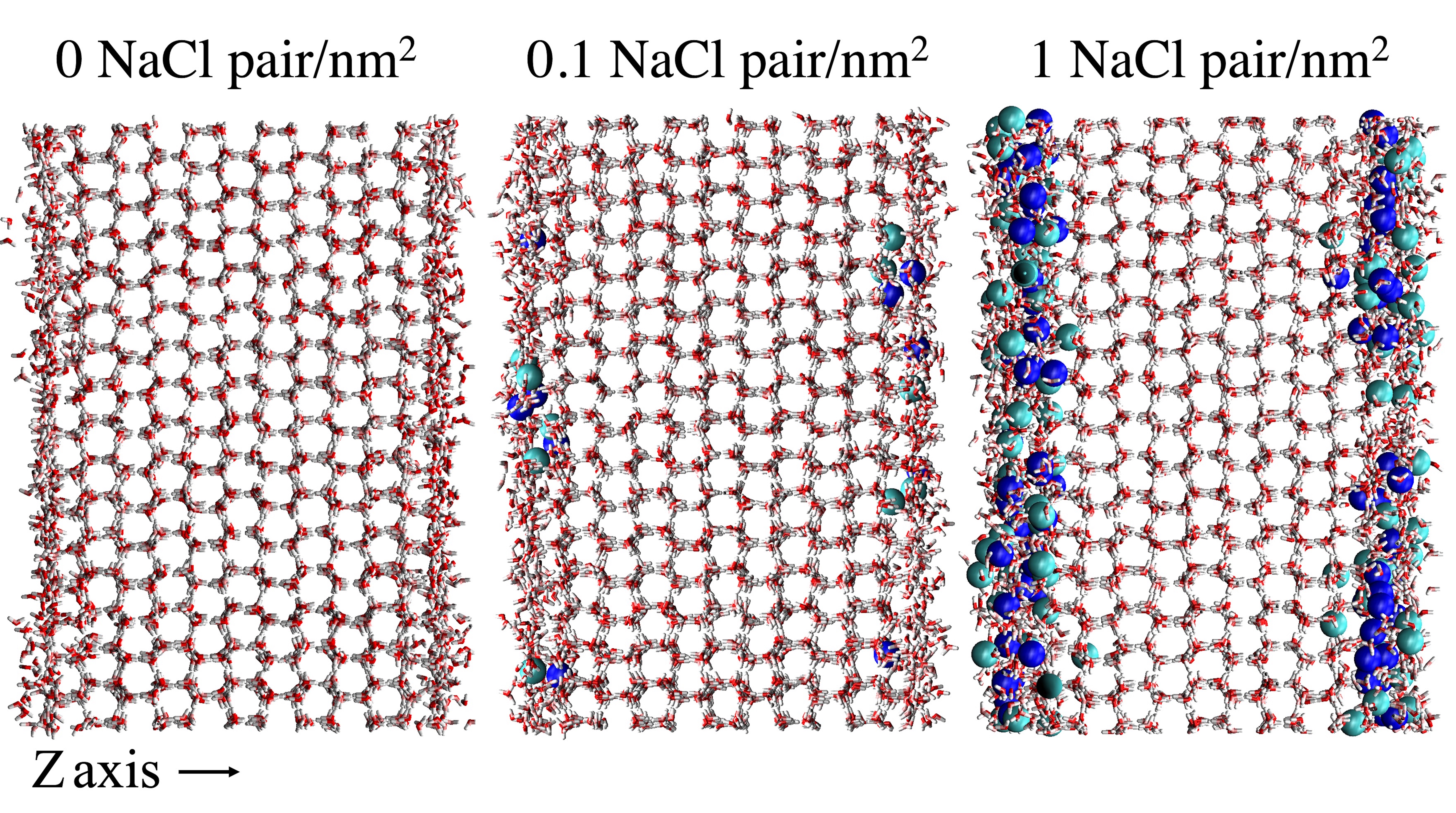}
    \caption{Snapshots of the molecular systems with the three varying surface surface densities of NaCl at the basal surface at 220 K}
    \label{fig:snapshot}
\end{figure}


\section{Results and Discussion}

To provide a microscopic insight into the surface structure of ice as a function of temperature and surface density of ion pairs, we computed density profiles, layer-resolved radial distribution functions, crystalline order parameter and surface roughness. 
Whereas we have conducted MD simulations with both low and high surface densities of ion pairs, hereafter we mostly focus on the comparison between pristine and briny ice surfaces with a surface density of 1 Na$^+/$Cl$^-$ pair per nm$^2$, i.e. a mass concentration of 47 parts per thousand (ppt) (see, for example, Figure~\ref{fig:snapshot}).  
The surface structure of ice with a low surface density of ion pairs (0.1 Na$^+/$Cl$^-$ pair per nm$^2$ or 5 ppt) do not exhibit significant differences from pure ice and the results for this system are reported in the supporting information file (Figures S1, S2, S7, S9, S11).

Using the experimental cryoscopic constant of water, the freezing point depression of ice with a surface density of 1 NaCl pair/nm$^2$ of NaCl is 3 K. However, the TIP4P/2005 model slightly overestimates the value of the cryoscopic constant and the system melts at 245 K.\cite{conde_high_2017}
We verified this prediction simulating the direct melting of ice slabs with sodium chloride ion pairs deposited at their surfaces.\cite{vega_absence_2006} 
Hence we consider temperatures up to 240 K, corresponding to -9 K below the melting temperature $T_m$ of pure ice and -5 K below $T_m$ with ions. 

 \begin{figure}[t]
    \centering
    \includegraphics[width=8.6 cm]{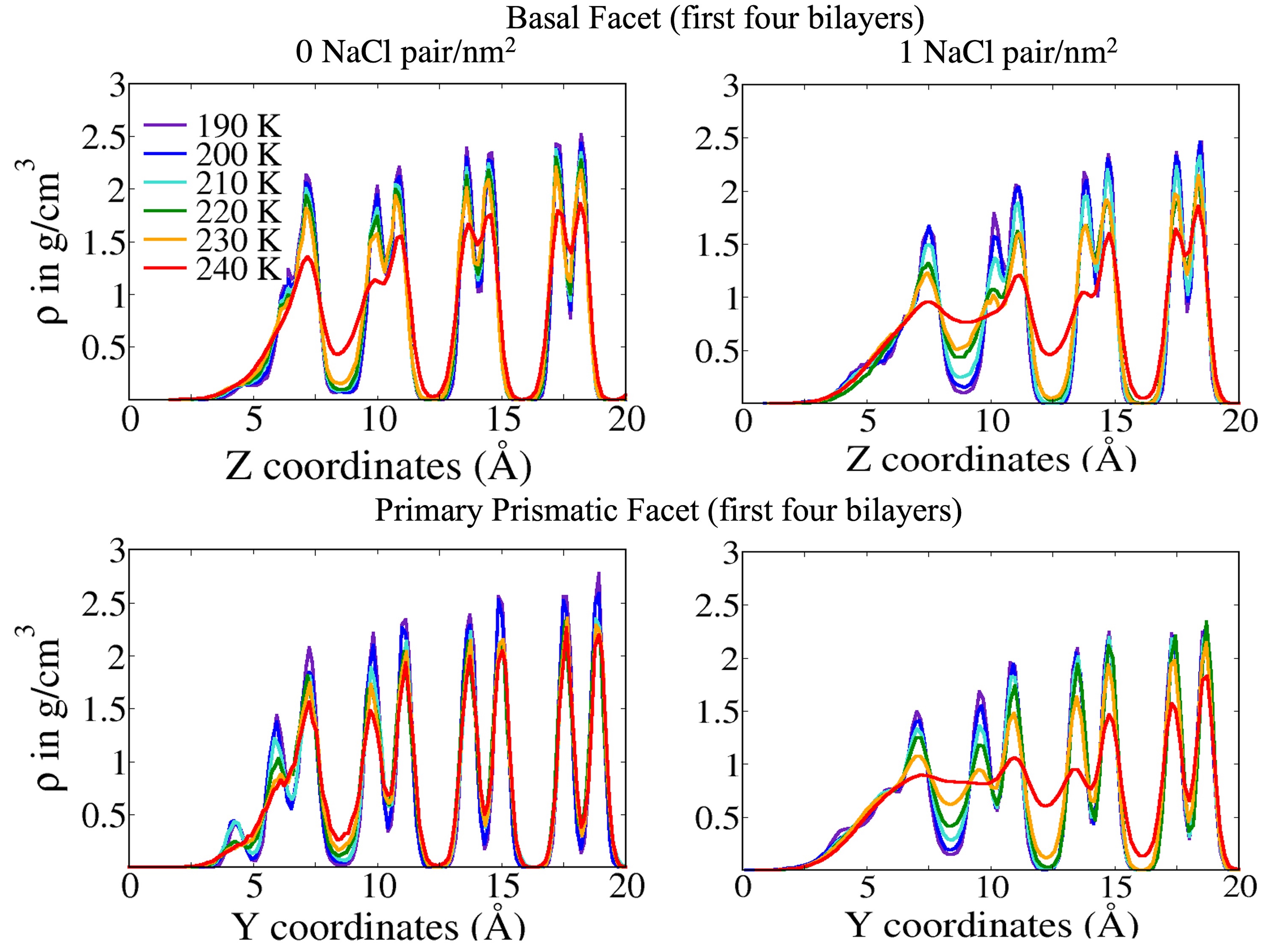}
    \caption{Oxygen atoms density profile as a function of temperature for the basal facet without (a) and with ions (b) and primary prismatic facet without (c) and with ions (d), averaged over 200 ns production runs. \textcolor{black}{Statistical uncertainties are of the order of the line widths.}}
    \label{fig:density}
\end{figure} 
\begin{figure*}[t]
    \centering
    \includegraphics[width=14 truecm]{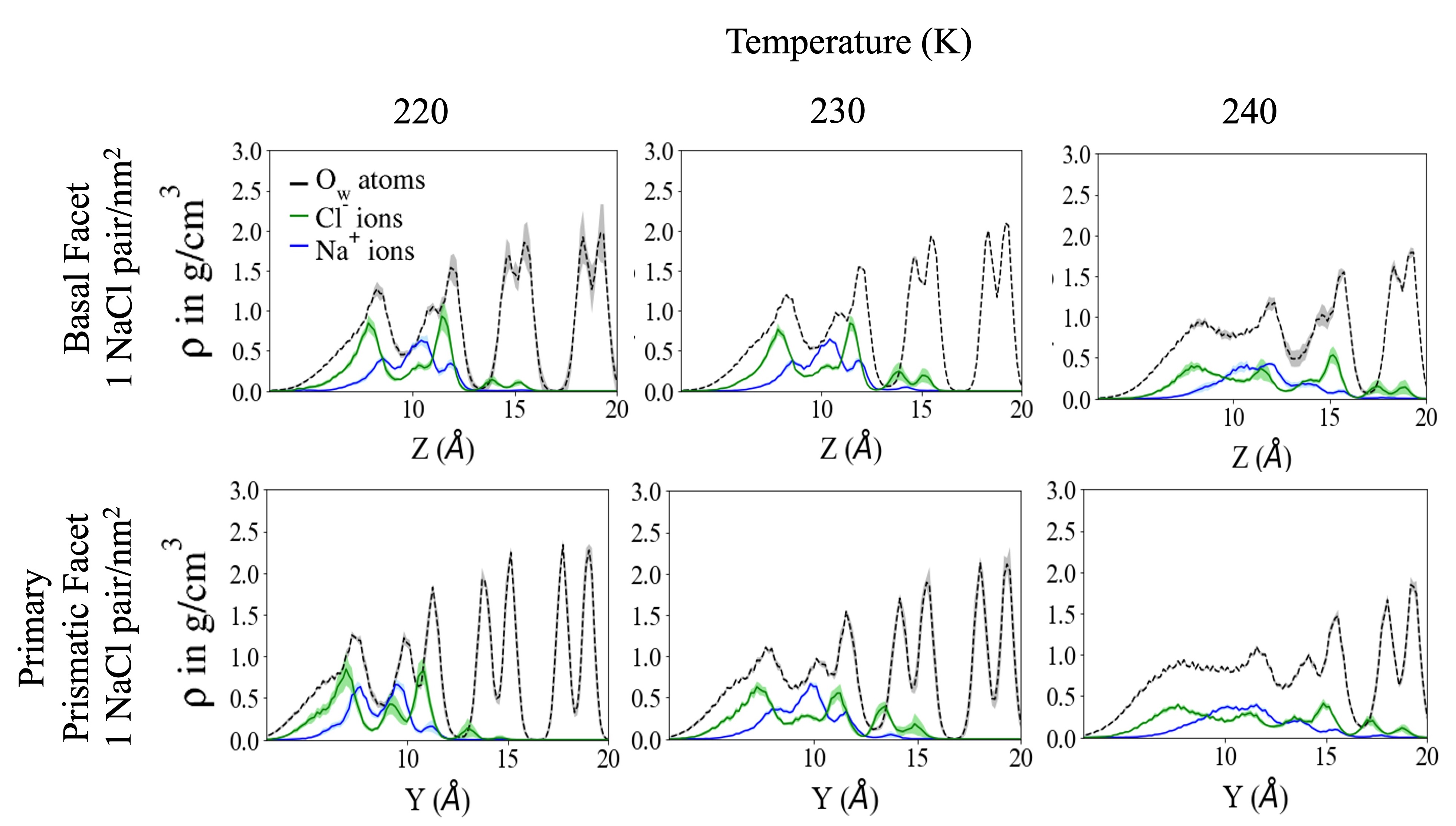}
    \caption{Cl\(^-\) and Na\(^+\) density profiles for the three highest temperatures sampled for both the basal and primary prismatic facet at the NaCl surface density of 1 NaCl pair/nm$^2$. Respective oxygen atom density profiles are included to show where the density of the ions is in relation to the bilayers. \textcolor{black}{Ion density is scaled by a factor of 5$\times$. 
    The shaded regions corresponds to the standard deviation of the density profiles from block statistical analysis.} 
    }
    \label{fig:iondensity}
\end{figure*}

\subsection{Density profiles}

Figure~\ref{fig:density} shows the density profile in the direction perpendicular to the plane of the surface. The air-ice interface is on the left side of the plots.
Both the basal and the primary prismatic surfaces consist of hydrogen-bonded bilayers. The density profiles show a double peak for each bilayer: a structural fingerprint, the absence of which provides a signature of disordering. As the temperature increases, the  outermost bilayers smear and their density peaks decrease, while the valley separating them from the layers underneath increases.
As observed in previous works, the outermost layer of pristine ice is already disordered at the lowest temperature considered (190 K). For example, for pristine ice the basal plane surface layer never exhibits the bilayer fingerprint (Figure~\ref{fig:density}a). The double peak is preserved in the first prism surface up to 230 K, but the presence of a third low-intensity peak at air-ice interface suggests the partial formation of an adlayer (Figure~\ref{fig:density}c). In the presence of ions the outermost bilayer of both the basal and the first prismatic surfaces are further smeared, and in the latter case, the bilayer structure and the adlayer-related peak are reduced to shoulders and eventually completely removed at 240 K (Figure~\ref{fig:density}b,d). 
The difference in the melting behaviour of the second and third bilayer is even starker. The sudden increase of the valley between the first and the second layer of the basal plane of pristine ice confirms the "rounded" bilayer-by-bilayer melting transition identified in previous works.\cite{llombart_rounded_2020} Conversely, in the presence of ions, the valley between the first and the second bilayer increases gradually, suggesting a continuous surface melting transition. 
The density profiles of the basal and the primary prismatic surfaces with ions (Fig.~\ref{fig:density}b,d) behave very similarly as a function of the temperature, suggesting that the ion pairs attenuate the structural differences between between these two surfaces.

\begin{figure}[h]
    \centering
    \includegraphics[width=8.6 cm]{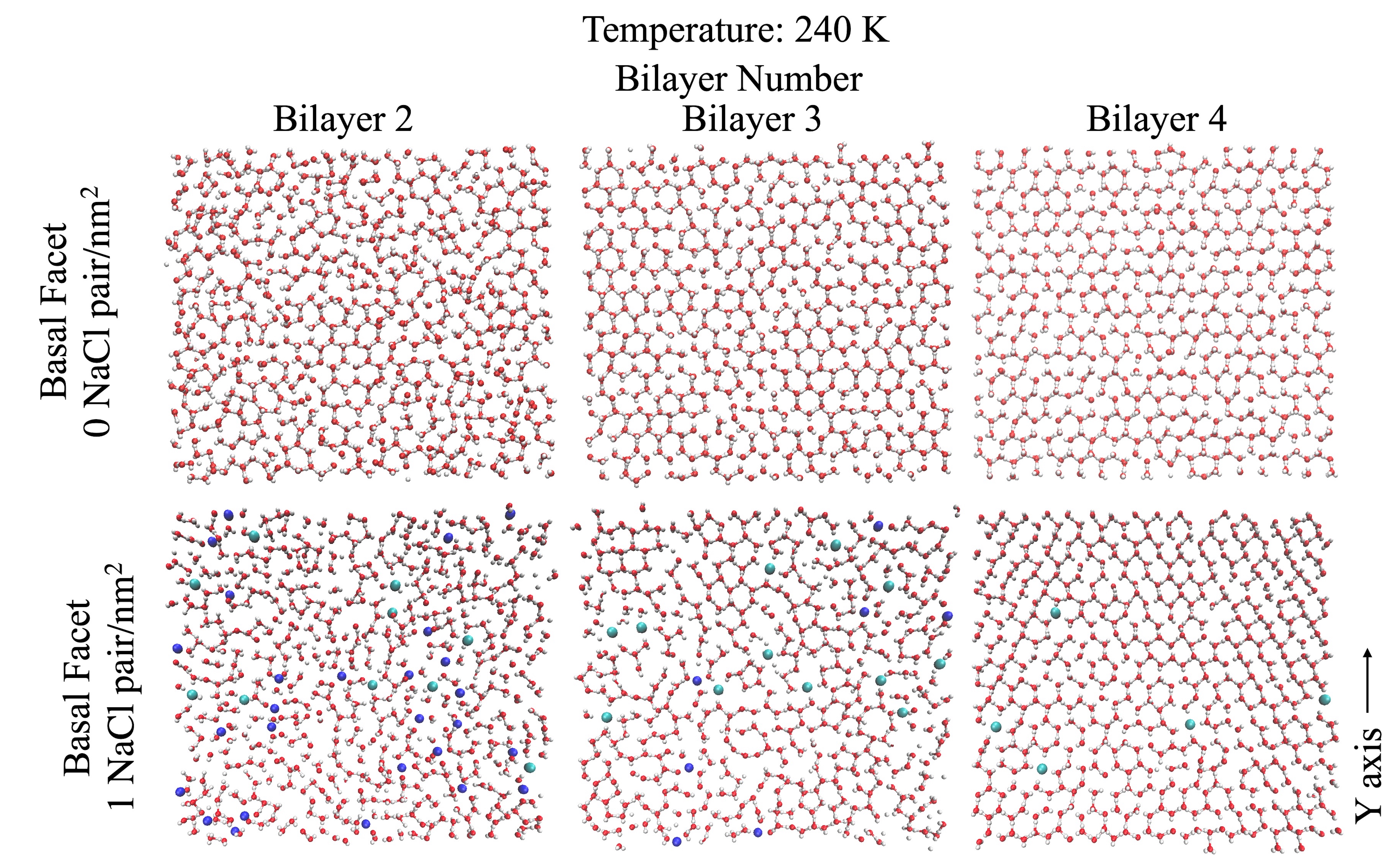}
    \caption{Snapshots from a top down view of the second, third, and fourth bilayers of the basal facet with and without NaCl at 240 K. Bonds shown are for a bond length cutoff 1.9 \(\AA\) to display the crystalline structure (or lack there of) in each layer. Na$^+$ ions are represented by blue spheres and Cl$^-$ ions are represented by green spheres.}
    \label{fig:layer snapshots}
\end{figure} 
The Na$^+/$Cl$^-$ density profiles (Figure \ref{fig:iondensity}) shows how ions penetrate in the subsurface layers as the temperature is increased from 220 to 240 K. 
At all temperatures Cl$^-$ is slightly in excess in the outermost part of the QLL, in accordance with the surface propensity of halides observed in previous experiments and simulations of aqueous solutions.\cite{piatkowski_extreme_2014, knipping_experiments_2000, ghosal_electron_2005, jungwirth_specific_2006, otten_elucidating_2012, tobias_simulation_2013}. 
\textcolor{black}{We have verified that our modeling framework produces consistent results with previous studies on the surface propensity of anions in aqueous solutions, even if the Madrid model is non polarisable. The atomic density profiles computed for a slab of aqueous NaCl solution at 248 K exhibit a slight accumulation of Cl near the surface and a depletion of Na in agreement with former MD simulations with polarisable water models (Figure S3).\cite{jungwirth_specific_2006, tobias_simulation_2013}}
%
%
On ice, the accumulation of negative charge at the surface causes a peak of Na$^+$ mass density in the first subsurface layer. Chloride is then prevalent in the layers underneath at the interface between QLL and crystalline ice. 
The density of chloride ions in the third bilayer increases gradually as the temperature rises from 220 to 230 K, and at 240 K Cl$^-$ penetrates into the fourth bilayer. This trend is the same regardless of surface orientation. 
\textcolor{black}{To shed light on the behaviour of ions in the QLL, we computed the charge density profiles and compared them with the particle densities of all four species in the system (O, H, Cl, Na) in Figure S4. 
The peaks of the density of Cl$^-$ align with the peaks of the density of H which have a positive partial charge, thus partially neutralizing the positive peaks of the charge density profile. Similarly, Na$^+$ tends to compensate the negative charge in correspondence to the peaks of the oxygen density. The overall effect is a rounding of the charge density features, which is also the fingerprint of the onset of disorder. 
}

\subsection{Structure of the quasi liquid layer}


\begin{figure*}[tb]
    \centering
    \includegraphics[width=17.5cm]{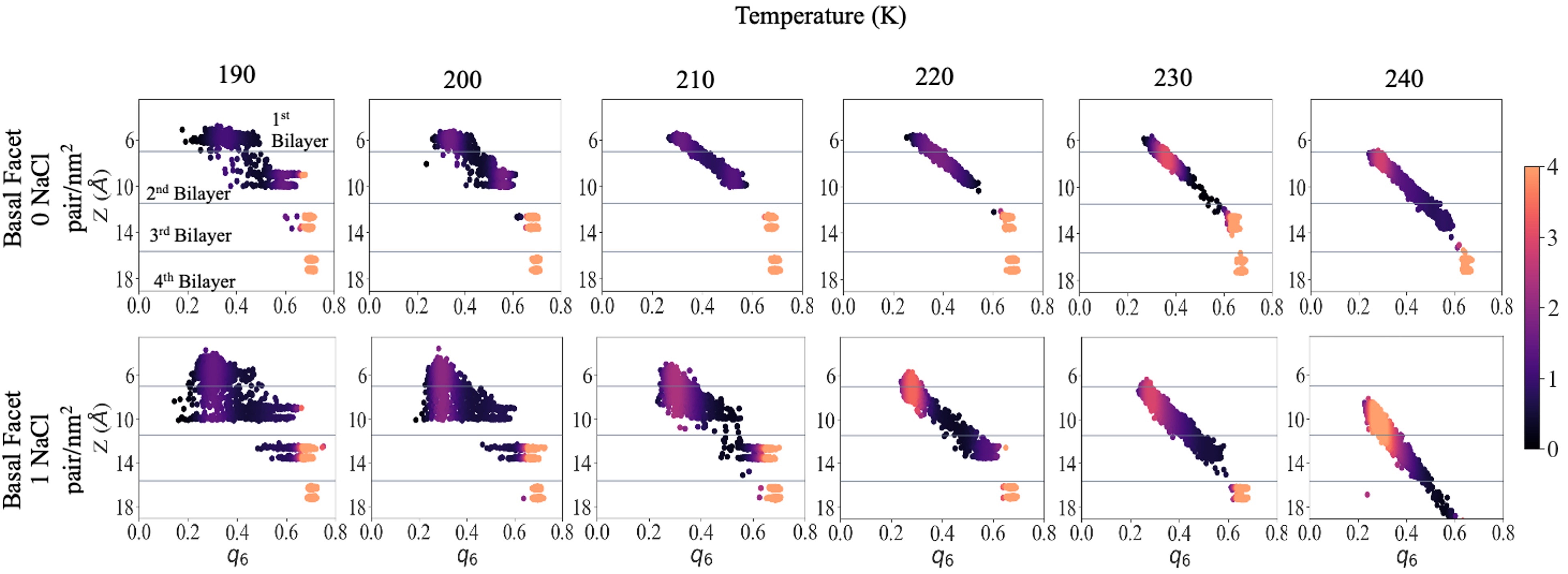}
    \caption{Local Steinhardt's order parameter $q_6$ density map of the first four bilayers at each sampled temperature, for the basal facet without and with ions. Averaged over 200 ns. The scatter plots are colored by the calculated Gaussian kernel density estimation for the q6 values, the scales values are arbitrary units but consistent for every density map. }
    \label{fig:order}
\end{figure*} 

Looking further into the structural changes induced by ion pairs, Figure~\ref{fig:layer snapshots} shows top-down snapshots  of the second, third and fourth bilayers of the basal facet at 240 K. Although the QLL is dynamic and diffusive, these representative snapshots allow us to interpret the molecular mechanism of ice premelting. 
At the highest temperature (240 K), we see that for pristine ice (Fig.~\ref{fig:layer snapshots} top panels) the third and fourth bilayers retain the crystalline structure of hexagonal ice, even though the third bilayer exhibits topological defects.\cite{donadio_topological_2005} The second bilayer is much more disordered but it still features areas with crystalline hexagonal structures. 
For the systems with 1 NaCl pair/nm$^2$ (Fig.~\ref{fig:layer snapshots} bottom panels), the second bilayer is completely disordered and contains a high surface density of ions -- mostly Na$^+$ -- as also shown by the ion density profiles. 
The third bilayer is mostly disordered due to the presence of ions, mostly chloride, and, similarly to the second bilayer of pristine ice, it exhibits crystalline patches. 
The fourth bilayer is mostly crystalline, with defects similar to those observed in the third bilayer of pristine ice. However, it embeds chloride ions that replace water in their crystallographic positions. As chloride is easily embedded into the ice lattice, when the temperature increases Cl$^-$ ions gradually penetrate into the deeper bilayers.\cite{smith_free_2005, vrbka_brine_2005} 
The negative ions cause a reorientation of the hydrogen bonding network around the crystallographic sites where they replace a water molecule. Additionally the preferred embedding of anions at the ice/QLL interfaces engenders a net charge that favours the formation of defects and destabilises the interfacial bilayer.\cite{kim_effect_2008} Previous work \cite{watkins_large_2011} found that the first two bilayers of pristine ice are more weakly bound and more susceptible to premelting than those deeper in the ice slab. The snapshots in Figure~\ref{fig:layer snapshots} suggest that the anions embedded in as far as the third and fourth bilayers destabilise the crystalline layer leading to a smoother premelting transition.

Characteristic snapshots of both the basal and the primary prismatic surfaces at different temperatures are reported in Figure~S5 for the second bilayer of pristine ice and in Figure~S6 for third bilayer of ice with 1 NaCl pair/nm$^2$. The comparison between these two sets of figures suggests that there is a close correspondence in the behaviour of the second bilayer of pristine ice to the third bilayer of briny ice across the temperature range explored. In the latter system the inclusion of chloride marks the onset of the premelting transition, but extended disordering occurs only when sodium penetrates in the bilayer. 
Merging the information from the density profiles and Figures~\ref{fig:layer snapshots}, S5, and S6 indicates that the presence of a high surface density of Na$^+/$Cl$^-$ ion pairs  not only shifts the melting temperature but it also systematically increases the thickness of the QLL by a full bilayer. The structural disordering of subsurface layers corresponds to the incorporation of Na$^+$ and Cl$^-$ ions, with the latter penetrating in the deeper layers earlier than the former.

The temperature evolution of the layer-resolved RDFs  of the surface and subsurface layers (Figure S7) contributes to support the hypothesis that the basal and primary prismatic surfaces, which have distinct premelting behaviour in freshwater ice, melt in a very similar way in the presence of ions and the premelting transition evolves more gradually as a function of temperature. 
%

%



\subsection{Local order parameter distribution}

To shed light on the disordering of the premelted ice layer as a function of temperature we calculated the local {\it q$_6$} Steinhardt order parameter\cite{steinhardt_bond-orientational_1983,ten_wolde_numerical_1996,reinhardt_local_2012} as implemented in former nucleation studies.\cite{moroni_interplay_2005, sosso_crystal_2016} 
Figure~\ref{fig:order} presents the density maps of the {\it q$_6$} order parameter of the first four bilayers of the basal facet at different temperatures. The density maps of the primary prismatic facet and of the basal facet at low surface density of ions are presented in Figure~S8 and S9 in the SI. 
For reference, Figure~S10 reports the $q_6$ distribution for a water slab at 240 K. The latter is between 0.2 and 0.3 and does not depend on the vicinity to the surface. 
Figure~\ref{fig:order} shows that for pristine ice at 190 K there is a sharp distinction in the q$_6$ values among the layers, which at this temperature remain well separated. The surface layer is disordered but its $q_6$ distribution remains on the right of the liquid range. The subsurface layer presents a broad but mostly crystalline $q_6$ distribution and the bottom two layers can serve as reference for crystalline ice. 
With high surface density of Na$^+/$Cl$^-$ ion pairs the first and the second bilayer are not clearly distinct and the $q_6$ distribution spans the whole range from the liquid to the ice reference. In particular a substantial part of the second bilayer has $q_6$ values corresponding to the liquid phase.
The third bilayer already shows the onset of  disordering  with a tail in the $q_6$ distribution extending below 0.5.

As the temperature is increased to 210 K, the first two layers of pristine ice form a continuous distribution of $q_6$ with a marked dependence on the z-coordinate. The $q_6$ distribution does not change significantly up to 230 K, except for a the shift due to the thermal motion of the water molecules and the onset of dynamic exchange of molecules between the second and the third bilayer. 
In the system with Na$^+/$Cl$^-$ ion pairs at 200 K and 210 K the first and second bilayers form a single  triangular $q_6$ distribution centred at $\sim 0.3$ extending to 0.6 for the subsurface layer. 
At 220 K and above there is a single $q_6$ distribution that encompasses the top three bilayers and strongly depends on the z-coordinate. The $q_6$ distribution for the outermost part of the QLL is indistinguishable from liquid water, whereas ordering gradually increases with depth in the subsurface region. 

At 240 K the $q_6$ of the third bilayer of pristine ice shifts to the left of the ice reference, indicating the onset of disordering. 
In the system with Na$^+/$Cl$^-$ ion pairs the $q_6$ distribution encompasses also the fourth bilayer, which retains values of the order parameters close to the ice reference.
Similar trends can be observed by comparing the primary prismatic facet with and without 1 NaCl pair/nm$^2$ (Figure S6).

The $q_6$ maps confirm that the thickness of the QLL increases by about one layer when ion pairs are deposited at high surface density. Remarkably, ions modify the local structure of the QLL weakening the regular template induced by the crystalline subsurface layers, so that the surface has the same distribution of the local order parameter as liquid water. 
This observation agrees with the conclusions of experiments showing that the premelting layer of seawater ice has the same vibrational properties as liquid brine.\cite{kahan_pinch_2014} 
As the ice template is weaker, the differences in the structure and the premelting behaviour between the basal and the primary prismatic surfaces are also weakened for the deeper subsurface layers and vanish for the outermost layers.


\subsection{Surface roughness}

The roughness of ice surfaces affects the adsorption of reactive species and determines the crystal growth rates of facets with different orientations, thus controlling the shape of crystals formed by condensation from vapour.\cite{libbrecht_physical_2017} 
The roughness of a surface is characterised by its \textcolor{black}{structure function} $H(R)^2$ defined as:
\begin{equation}
     H(R)^2  =  \langle (\Delta h(0) - \Delta h(R))^2 \rangle,
     \label{eq:H}
\end{equation}
where $h(R)$ is the height of the surface at the planar distance $R=|(R_x,R_y)|$ with respect to the plane origin, \textcolor{black}{$\Delta h(0)$ = $h(0)$ - $\langle h(R) \rangle$,} and $\langle\ldots \rangle$ indicates an ensemble average that runs over the frames of the MD trajectory and the position of the origin.
\textcolor{black}{We define the surface by identifying the oxygen atoms, and sodium and chloride ions at the air/ice interface as those accessible to a spherical probe with a 3.5~\AA~radius.\cite{sega_pytim_2018}}
$H(R)$ in Eq.~\ref{eq:H} can be rewritten in terms of the normalised height-height correlation function $C(R) = \langle \Delta h(0)h(R) \rangle /\langle \Delta h(0)^2 \rangle $ and of the surface roughness $\sigma = \sqrt{\langle \Delta h(0)^2 \rangle} $ as: 
\begin{equation} \label{eq:Correl}
  H(R)^2  =  2\sigma^2[1 - C(R)].
\end{equation} 

\textcolor{black}{
A disordered surface is defined ``flat" when it does not have long-range correlations. In this case, which is common for solid surfaces, $C(R)$ decays exponentially with a characteristic length $\lambda$ and the surface roughness squared $\sigma^2$ is the large-$R$ limit of $H(R)^2$.\cite{rommelse_preroughening_1987}
The surface of a liquid, instead, is ``rough" and $H(R)$ diverges as $\log(R)$. The logarithmic divergence of the interfacial width at the air/liquid interface of water is due to the presence of capillary waves.\cite{ismail_capillary_2006} 
Previous studies using the TIP4P/ice water model showed that, even in the presence of a premelting layer, the air-ice interface can be considered flat up to few degrees below the melting point, and it undergoes a sequence of phase transitions.\cite{benet_premelting-induced_2016, llombart_surface_2020} Here we revise these findings with the TIP4P/2005 water model and we address the effect of ions. To classify surface structures and transitions, we use the same nomenclature used in previous works addressing surface roughening and premelting.\cite{jagla_surface-melting-induced_1999, llombart_surface_2020}
}

\begin{figure}[tb]
    \centering
    \includegraphics[width=8.5 cm]{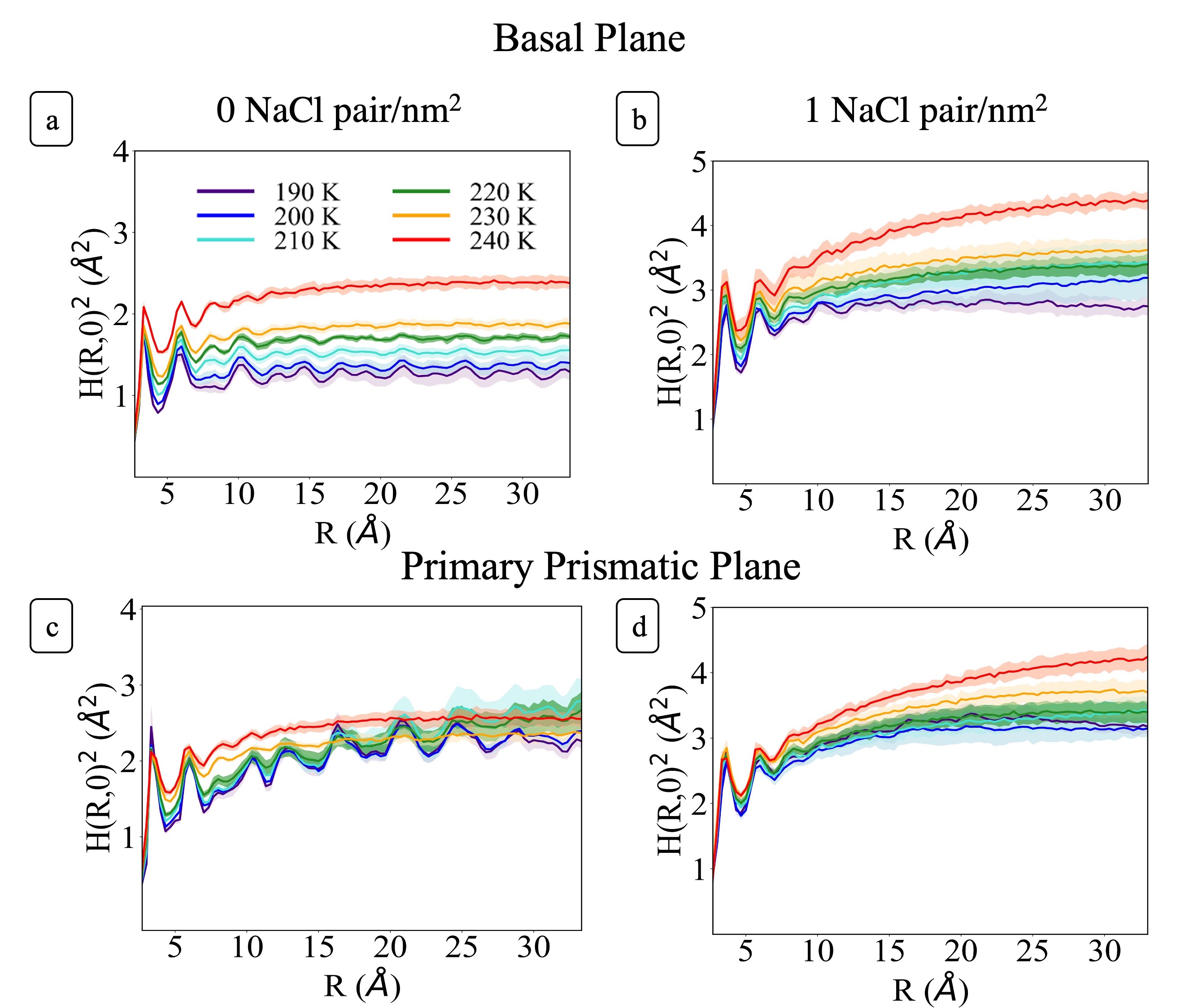}
    \caption{Structure function as a function of temperature \textcolor{black}{ of the basal facet of pristine ice (a) and of ice with sodium chloride ions (b), and of the primary prismatic facet of pristine ice (c) and of ice with sodium chloride ions (d).
    The shaded regions corresponds to the standard deviation of the structure functions from block statistical analysis.}
    }
    \label{fig:heightC}
\end{figure} 

\begin{figure*}[bt]
    \centering
    \includegraphics[width=17.5 cm]{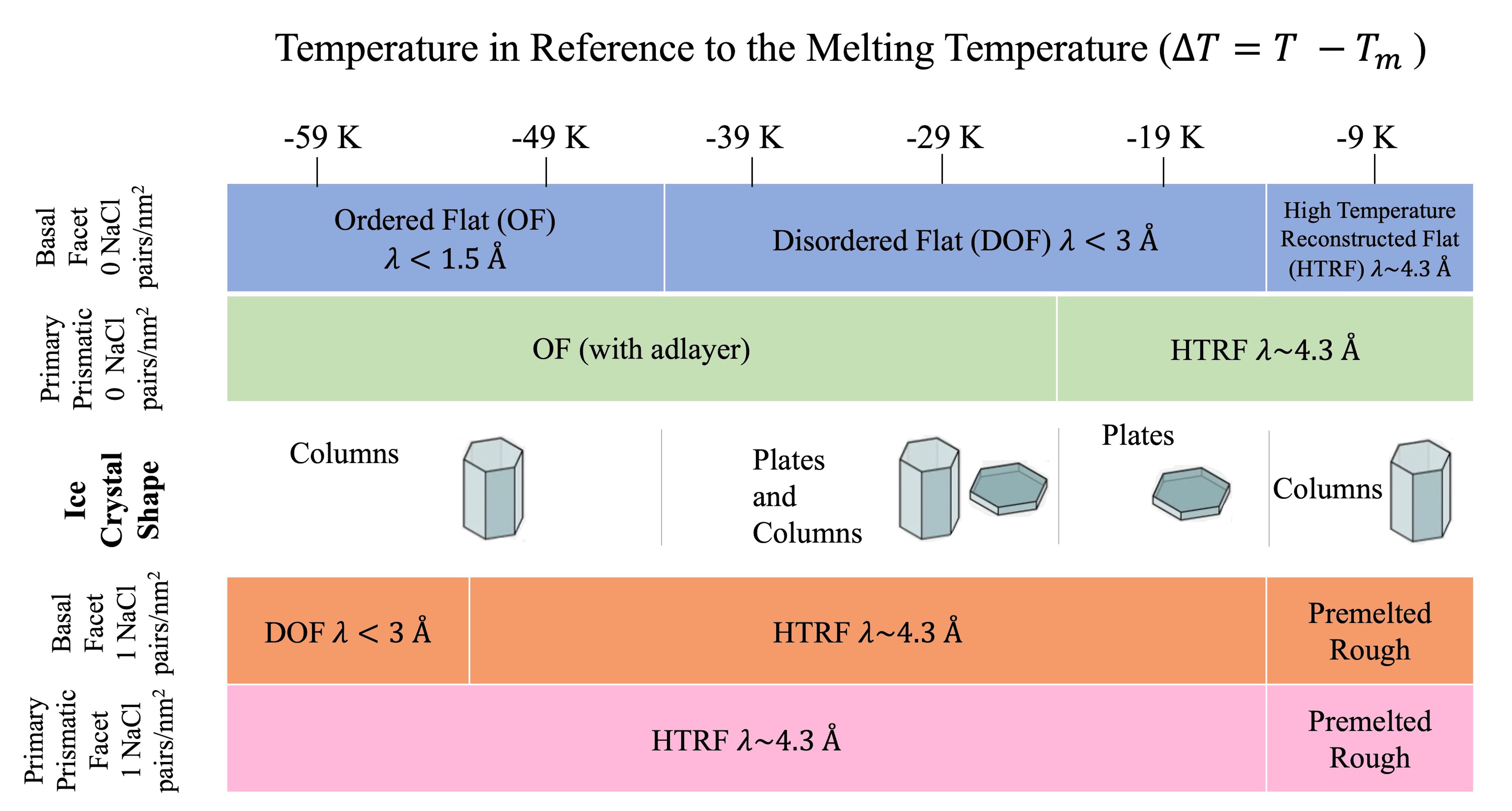}
    \caption{Schematic of ice crystal growth at low supersaturation in relation to the correlation length values obtained from Figure~\ref{fig:heightC}. Ice crystal shape temperature dependent transitions obtained from ref.\cite{slater_surface_2019}}
    \label{fig:crystalshape}
\end{figure*} 

Figure~\ref{fig:heightC} shows $H(R)^2$ at each temperature for the basal and primary prismatic facets with and without  Na$^+$Cl$^-$ pairs. The curves show \textcolor{black}{two} distinct peaks for $R<7$~\AA\ which are characteristic of the short-range hydrogen bonding environment. 
At larger $R$ the functions differ from one another with a marked dependence on ions surface density, orientation, and temperature.
$\sigma$, $\lambda$, and $H(R)^2$  for the two different orientations of pristine ice are rather distinct.  
For the basal plane (Figure~\ref{fig:heightC}a)  from 190 to 230~K, $H(R)^2$ reaches its asymptotic roughness very rapidly and the value of $\sigma^2$ remains below 2~\AA$^2$, indicating that the surface is flat. 
At temperatures below 210~K the surface exhibits a very short correlation length $<$1.5~\AA.  $H(R)^2$ retains periodic fluctuations, which are the signature of both the non-diffusive character of the system and residual hexagonal patterning. Hence, we classify this phase as \textcolor{black}{ {\sl ordered flat} (OF).} 
The periodic features begin to fade for $R \gtrsim 10$~\AA\  starting at 210~K and $\lambda$ increases to $\sim 3$~\AA. This indicates a surface phase transition from OF to {\sl disordered flat} (DOF), which is accompanied by the onset of surface diffusion. 
At 240~K $H(R)^2$ retains similar characteristics as the lower temperatures but the roughness jumps beyond 2~\AA, and  \(\lambda\) reaches a value of approximately 4.3 \AA, indicating a second structural transition to a {\sl high temperature reconstructed flat phase} (HTRF). 

Conversely, the $H(R)^2$ of the primary prismatic surface (Figure~\ref{fig:heightC}c) displays large periodic fluctuations from 190 to 220 K. The convergence toward the asymptotic values is slower, and in two cases (210 and 220~K) the correlation length is comparable or larger than the maximum radius of a sphere that can be enclosed in the simulation cell.
\textcolor{black}{The long-range fluctuations indicate that this surface  retains the fingerprint of crystalline ordering, although the structure is affected by structural defects that lead to the formation of an adlayer of molecules (see also Figure~\ref{fig:density}). Hence, we dubbed this surface {\sl ordered flat} (OR).}
The roughness parameter $\sigma^2$ is non-monotonic with the temperature. It reaches a maximum at 210~K, drops significantly at 230 K and increases again at 240~K. 
At 230~K the periodic features of $H(R)^2$ vanish indicating a structural transition 
to HTRF with \(\lambda\) values of approximately 4.3 \AA\ and roughness similar to that of the basal plane at 240 K. 

\begin{figure}[tb]
    \centering
     \includegraphics[width=7.5cm]{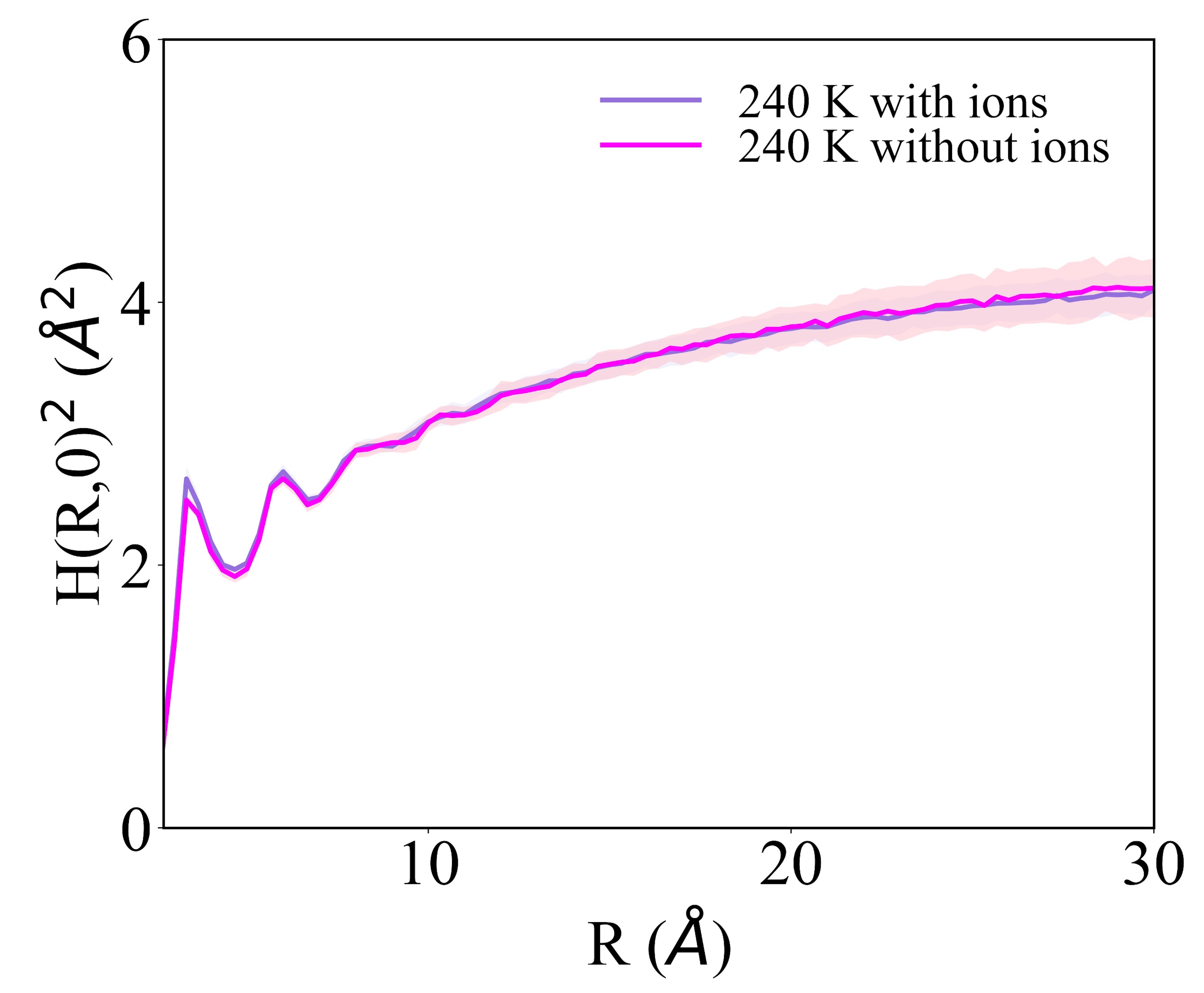}
     \caption{Structure function of the surface of water and of aqueous NaCl solution with a concentration of 45 ppt at 240 K, computed for a slab of 6144 water molecules. The two structure functions overlap, and exhibit logarithmic divergence. 
    }
    \label{fig:waterslab}
\end{figure}

\textcolor{black}{The sequence of preroughening transitions of pristine ice surfaces is similar to those observed in Ref.~\cite{llombart_surface_2020} and correlate with changes in the growth behaviour in the Nakaya diagram, which classifies the shape of ice crystals growing from vapour deposition at different temperatures and humidity (Figure~\ref{fig:crystalshape}).\cite{Nakaya, libbrecht_physics_2005, libbrecht_physical_2017}}
The transition from OF to DOF of the basal surface, at about $T_m -40$~K, corresponds to the change from columnar ice crystals to plates at low vapour pressure. The transition to HTRF for the primary prismatic surface at about $T_m -25$~K corresponds to the preference for plates at any vapour pressure. Finally, the transition to HTRF of the basal surface at $T_m -15$~K correlates with the occurrence of columnar crystals. 


$H(R)^2$ for the systems with 1 NaCl pair/nm$^2$ (Figure~\ref{fig:heightC}bd) indicates that these surfaces are much rougher than the corresponding pristine ice facets. Additionally, even at the lower temperatures, no features arising from a periodic pattern are present, thus suggesting that solvated Na$^+/$Cl$^-$ ion pairs disrupt long-range order at the air-ice interface. 
\textcolor{black}{The primary prismatic surface is generally rougher than the basal surface up to 230 K. At 190 K, the basal surface with ions has a similar structure function and correlation length to that of pristine ice in the DOF phase, which occurs at higher temperature. 
At temperatures between 200 and 230 K, the basal and primary prismatic surfaces with ions have the same features and correlation lengths, and can be classified as HTRF. Above 230 K both surfaces undergo a roughening transition that leads to a logarithmically divergent $H(R)$. The structure function of this phase, named premelted rough, is indistinguishable from that of a liquid water slab at a similar temperature (Figure \ref{fig:waterslab}). These results suggests that the surface structure of ice with a high density of NaCl is characterized by the occurrence capillary waves. 
Whereas previous works suggested that anions pin equilibrium capillary waves at the surface of either water or ice,\cite{niblett_ion_2021, otten_elucidating_2012} we do not observe this effect in our simulations. Figure~\ref{fig:waterslab} shows that, in spite of the surface propensity of Cl$^-$, neither the surface of a liquid slab is affected by the dissolved NaCl salt, nor we observe the pinning of capillary waves in the premelting layer.}
 
Above 200 K ($T-T_m \sim -40$ K) the surface roughening transitions occur at the same temperatures for both facets, thus affecting the shape of growing ice crystals. We indeed expect that the presence of ion pairs at typical atmospheric temperatures would level the growth rates of the two low-index surfaces and lead to the formation isotropic crystals.

\section{Conclusions}
We have characterised the structure and melting behaviour of both the basal and primary prismatic surfaces of ice \(I_h\) in the presence of sodium chloride at both high and low surface density. 
We find that the addition of ions in low surface density does not change the structural properties of the QLL with respect to freshwater ice. 
Conversely, high surface density of ion pairs corresponding to seawater ice conditions leads to a significant change in QLL thickness, structure and premelting behaviour. 
Na$^+/$Cl$^-$ ion pairs disrupt the residual ordering of QLL and increase its thickness by at least one bilayer at all the temperatures considered. Surface premelting becomes more gradual in the presence of ions and the differences in ordering and premelting behaviour between the two surfaces are erased. 
Close to the melting temperature \textcolor{black}{(T - $T_m\sim -10$~K)} the QLL with ions exhibits the same structural features as a water slab, thus providing the same solvation environment as a liquid surface.\cite{kahan_pinch_2014}

The changes induced by Na$^+/$Cl$^-$ ion pairs at the molecular level lead to major chances in the surface structural transitions as a function of temperature, erasing the differences between the two facets. This effect may impact the growth rate and the shape of ice crystals growing from the vapour phase in clouds containing significant amounts of sodium chloride aerosols. 
The observed changes in surface roughness upon ion adsorption may also impact the light scattering properties of ice and snow. The observation that adsorbed ions increase surface roughness, knowing that rougher surfaces decrease albedo, would explain why sea ice has a lower albedo than freshwater snow.

In general terms, the key implication of our study is the crucial need to consider the surface density of Na$^+/$Cl$^-$ present when modelling ice surfaces, given its ability at high surface density to alter the surface structure, thus affecting mechanistic properties such as glacier sliding,\cite{cuffey_deformation_2000} and chemistry at the ice surface.\cite{shepherd_quasi-liquid_2012,bartels-rausch_review_2014,abbatt_interactions_2003}

\section*{Conflicts of interest}
There are no conflicts to declare.

\section*{Acknowledgements}
We are grateful to Luis G. MacDowell for useful discussions and the critical reading of the manuscript. 
This work is supported by the National Science Foundation under Grant No. 1806210 and Grant No. 2053235. 



\balance


\bibliography{rsc_updated5.bib}

\bibliographystyle{rsc} 

\end{document}